\documentstyle[12pt,aasms]{article}
\tightenlines

\begin{document}

\title{The Pulsating Pre--White Dwarf Star PG0122+200}

\author{M. Sean O'Brien \altaffilmark{1}, 
J. Christopher Clemens \altaffilmark{1,2},
Steven D. Kawaler \altaffilmark{1,3}, and
Benjamin T. Dehner \altaffilmark{1}}
\altaffiltext{1}{Department of Physics and Astronomy, Iowa State University, 
Ames, IA  50011  USA; msobrien@iastate.edu}
\altaffiltext{2}{Hubble Fellow}
\altaffiltext{3}{Visiting Astronomer, Institute of Astronomy, Cambridge, UK}

\begin{abstract}

We present an analysis of single--site time--series photometry of the pulsating 
pre--white dwarf PG~0122+200.  We show the pulsations are consistent with
a pattern of modes equally spaced in period; both the observed period range 
and spacing confirm that PG~0122 is a $g$--mode pulsator.  PG~0122 shows a 
pattern similar to that seen in multi--site observations of PG~2131+066 and 
PG~1159--035.  The measured period spacing, combined with the spectroscopic 
temperature, constrain the stellar mass much more precisely than the published 
measurement of its surface gravity.  Based on stellar models, the mass of 
PG~0122 falls in the range 0.66--0.72~$M_{\odot}$.  Fine structure in the 
power spectrum indicates that PG~0122 rotates once every 1.6 days.  Future 
multi--site observation (e.g., using the Whole Earth Telescope) should increase 
the precision of these results and reveal detailed information on the internal 
structure of this variable pre--white dwarf star.

\end{abstract}


\section{Introduction}

The PG~1159 stars represent the ephemeral penultimate stage in the life of 
low-- and intermediate--mass stars.  Gaining knowledge of their properties 
provides insight into their predecessors---central stars of planetary
nebulae and AGB stars---and their descendants---the white dwarfs.  The 
pulsations exhibited by some PG~1159 stars help us in this endeavor.  With 
the tools of asteroseismology in hand, the observed frequencies can be used 
to determine the total stellar mass and surface layer mass of the 
compositionally stratified pre--white dwarfs.  This then constrains their 
genealogy and structure.

The variable PG~1159 stars (GW Vir stars) are non--radial $g$--mode pulsators. 
Theory suggests---and observations show---that for such high surface gravity 
objects (log $g \sim 7$), the power spectra should be rich but essentially
well ordered: we expect to see patterns of modes equally spaced in period.  
High radial overtone $(n \gg 1)$ modes of low spherical harmonic index $\ell$ 
can show multiplet structure; rotation can split each one into $2\ell+1$ peaks 
in the power spectrum.  Other effects, such as a stellar magnetic field, 
cause frequency splitting about the $m = 0$ mode in a given multiplet.  This 
splitting can be asymmetric depending on the field geometry.  

Despite the potential complexity, only modes with low $\ell$ have been 
identified in GW Vir stars.  Thus the power spectrum is not necessarily 
complex beyond comprehension.  Kawaler \& Bradley (1994) showed that the 
average period spacing between modes of the same $\ell$ and consecutive $n$ 
depends primarily on the total stellar mass, with lesser dependence on 
luminosity and only a very slight dependence on composition.  Periodic 
deviations from mean period spacing can reveal the existence of a composition 
interface (Kawaler 1988; c.f. Kawaler \& Bradley, 1994).  Determination of 
the values of $n$, $\ell$, and the azimuthal quantum number $m$ therefore 
reveals a wealth of information about the structure of the star 
(c.f. Winget et al. 1991 \& Bradley 1994).  

The task of decoding the power spectrum of a real star is complicated 
because not all possible modes are necessarily present.  Also, the periodic 
intervention of the Earth between the telescope and the star introduces 
aliases into the frequency spectrum which confuse mode identification.  The 
aliasing problem currently is addressed using the Whole Earth Telescope (WET,
Nather et al. 1990). A major success of WET came in 1989, when observers at 
telescopes around the globe obtained two weeks of almost continuous data on 
the class prototype, PG~1159--035 (hereafter PG~1159).  This minimized the 
aliasing and allowed Winget et al. (1991) to conclusively identify pulsation 
modes.  They found an unbroken sequence of almost twenty multiplets---each 
corresponding to $\ell=1$ and consecutive $n$---along with several $\ell=2$ 
modes.  The total stellar mass, the rotation rate, and the envelope mass were 
determined with unprecedented precision, showing that asteroseismological 
models could be used with confidence to study the PG~1159 pulsation 
phenomenon.  Recently, Kawaler et al. (1995) analyzed a second PG~1159 star, 
PG~2131+066, again using WET data. PG~2131 displayed a much simpler frequency 
spectrum:  only a few consecutive $\ell=1$ triplets, and no $\ell=2$ modes, 
were identified.  The data from WET observations of other PG~1159 stars are 
currently in analysis.

The star PG~0122+200 (BB Psc, $m_{b} = 16.13$) was identified as a member of 
the PG~1159 spectral class by Wesemael, Green, \& Liebert (1985).  Dreizler 
et al. (1995) report a surface gravity of log $g=7.5\pm0.5$, and an 
effective temperature of $75,500\pm5,000$~K, placing PG~0122 among the coolest 
PG~1159 stars.  Bond \& Grauer (1987) discovered variability in PG~0122, 
reporting a power spectrum dominated by variations at 402.3~s and 443.7~s.  
Hill, Winget, \& Nather (1987, hereafter HWN) observed the star in white light 
on four consecutive nights in late 1986 with the 2.1 m reflector at McDonald 
Observatory.  They tentatively identified eight pulsation modes between 300~s 
and 700~s, and suggested a mean period spacing of 16.4~s.  This implied a mass 
of $\sim0.7 M_{\odot}$, based on models developed by Kawaler (1987).  

At the time HWN published their observations, rotational splitting had not 
been resolved in the power spectrum of any of the hot degenerate stars, so that 
identification of the value of $\ell$ and $n$ for individual modes in these 
stars remained uncertain.  WET observations have since led to measurement of 
$\ell=1$ rotational splitting for PG~1159 ($\delta\nu = 4.22$ $\mu$Hz) and 
PG~2131 ($\delta\nu = 27.4$~$\mu$Hz), implying rotation rates (see equation~[3] 
below) of 1.38 days and 5.07 hours respectively.  All of the high--amplitude 
variation in both stars is attributed to $\ell=1$ triplets, with a series of 
low amplitude $\ell=2$ modes discovered in PG~1159. 

Armed with insight gained from these observations, we reanalyzed the HWN data.  
We hypothesized that the power spectrum of PG~0122 was comprised of $\ell=1$ 
rotationally split triplets.  If PG~0122 had the complex mode structure of 
PG~1159, this analysis---based on single--site data---would have failed.   
Luckily, PG~0122 is a simple star that is remarkably similar to PG~2131 in the 
sparseness of its frequency spectrum.  This allowed us to test our hypothesis 
without multi--site observations.  This paper describes our reanalysis of the 
HWN data.  In the next section, we describe the observations and their 
reduction.  Section~3 outlines our analysis, including our efforts to separate 
peaks from aliases and the evidence for equal period spacing in the power 
spectrum.  In section~4, we use the periods to constrain the physical 
properties of the star, and we conclude with section~5.

\section{Light Curves and Power Spectrum}

HWN obtained a total of 4,609 5--second and 6,137 10--second integrations over 
four consecutive nights (see Table 1). We reduced these data following the
procedures in Nather et al. (1990).  Figure~1 presents the data after summing 
to produce 40--second integrations.  Subsequent analysis was performed on the 
unsummed data.  Note the change in amplitude in the light curve from night to 
night, which was mentioned by HWN as possibly deriving from fine structure in 
the frequency spectrum of PG~0122 (i.e., beating).  

Figure~2 shows the power spectrum of the entire data set, out to the highest 
frequency showing significant power above noise.  The power spectrum was 
obtained by squaring the modulus of the Fourier transform (FT) of the light 
curve, resulting in a plot of modulation power (mp) versus frequency, where 
mp~=~($\Delta$I/I)$^{2}$, as discussed in Winget et al. (1994).  The low 
frequency power (below about 1000 $\mu$Hz) is due primarily to extinction and 
sky brightness variations in the atmosphere.  The decrease of power at zero 
frequency is a consequence of the data reduction procedure.

The dashed lines in Figure~2  indicate the power $P$ corresponding to different 
false alarm probabilities $FAP$($P$), calculated according to
\begin{equation}
FAP(P) = 1 - \left(1 - e^{-P/<P>}\right)^{N}
\end{equation}
where $N$ is the number of data points and $<$$P$$>$ is the average power,
calculated over the $N$/2 independent frequencies $f_{i}$ which characterize 
the data:
\begin{equation}
f_{i} = i/T~~~~~~~~~~~~~~~~i = 1,2,3,...,N/2.
\end{equation}
$T$ is the total time spanned by the data set.  $FAP$($P$) is the probability 
that noise will generate a peak of a given power at least once among the 
frequencies in equation~[2].  Though power spectra are usually presented 
over--resolved for clarity, calculating the FT at more than N/2 frequencies 
adds no information about the data, since the ``extra'' frequencies are not 
independent of those in equation~[2] (c.f. Horne \& Baliunas, 1986).  Unlike a 
statistical analysis which evaluates confidence by considering only how many 
standard deviations a peak stands above the local mean, the false alarm 
probability also accounts for the increasing probability that a peak of 
specific size will occur by chance as the number of independent data points 
increases.  For example, when an FT includes 25,000 independent frequencies, 
the chance that at least one noise peak will have three times the mean power 
is $\sim$99\%.  The false alarm probability provides a much more conservative 
and realistic criterion for judging the significance of a peak in the power 
spectrum.

We reiterate that we calculate $<$$P$$>$ over {\it all} the frequencies in 
equation~[2], from $f=1$/T to the Nyquist frequency N/2T.  Thus even though FAP 
analysis assumes normally distributed noise, we chose not to filter the data 
either to remove the low-frequency ``atmospheric noise'' (which is certainly 
{\it not} normally distributed) evident in Figure~2, or to remove the real 
stellar variations. As a result, our FAP levels are overly conservative for the 
frequency range we analyze, but we prefer this to the alternative which 
requires subjective evaluation of atmospheric noise.  Consequently, our FAPs 
should not be regarded as an absolute measure of peak significance, but as an 
objective, conservative guide that the reader can apply to the power spectrum 
of PG~0122.
 
Figure~3 shows the frequency region of primary power (excluding low frequency 
atmospheric variations) in the spectrum of PG~0122.  For comparison, the 
spectral window (on the lower right of Figure~3) depicts the transform of a 
single noise--free sinusoid, sampled at the same times as the actual data. 
This power spectrum shows seven groups of peaks, six of which correspond to 
periods identified in 1987.  We were unable to confirm the existence of 
significant variation at 435~s and 364~s (2300~$\mu$Hz and 2750~$\mu$Hz) 
reported by HWN.  There is evidence for slight excess power at these periods, 
but so close to noise that we did not include them in our analysis.  One 
additional peak, not listed in the 1987 paper, is clearly discernible in the 
transform, at 570~s (1755~$\mu$Hz).  

\section{Data Analysis}
\subsection{Frequency Identification}

Comparison to the spectral window shows that most frequency groups are more 
complex than the pattern formed by a single periodicity, indicating the 
possible presence of a companion.  Only the group near 1630~$\mu$Hz (and 
perhaps the one at 1755~$\mu$Hz) apparently represents the spectrum of a 
single sinusoid.  The other groups, and especially the region around 
2640~$\mu$Hz, are more complex than the window pattern.  In particular, the 
group at 2220~$\mu$Hz seems at first glance to represent the combination of 
two window--like peak groups, one of which is smaller in amplitude and offset 
from the other by a few $\mu$Hz in frequency.  

To separate closely spaced frequencies, we consecutively calculated linear 
least--squares fits for all eight identified frequency groups.  The data were 
then ``prewhitened'' at the largest peak in a region to look for lower 
amplitude variations.  This was accomplished by subtracting from the light 
curve, point by point, a sine wave of the frequency, amplitude and phase 
determined by the least--squares analysis of a given peak.  Smaller peaks 
identified in this way were then fitted with sine waves simultaneously with 
the larger peaks in the original data set.  The precision of frequency 
determinations for the previously known peaks was also greatly improved by 
this process. This procedure has proven highly successful in recovery of 
low--amplitude variations from WET data, where even the residual alias can 
swamp small peaks.  We describe below the details of this method as we applied 
it to the archival PG~0122 data.  Table~2 summarizes the frequencies, 
amplitudes, and mode identifications for the periodicities found in the light 
curve of PG~0122.   

In the process of prewhitening we encountered the inevitable problem of 
distinguishing real power from sampling aliases.  In every case, we applied the 
simple algorithm of choosing the largest (and therefore most statistically 
likely) peak as representing the real power.  Thus although our identification 
of peaks is not the only one consistent with the data, it is the most likely 
solution. In the next section we show that the rules we followed lead to a set 
of observed peaks with frequencies that are entirely consistent with 
rotationally split $\ell=1$ triplets equally spaced in period.

Starting with the region of greatest power (the group at 2500~$\mu$Hz) we 
assumed that the largest peak represented a true pulsation mode and we 
removed it.  Figure~4 shows the result.  The remaining power spectrum, 
depicted in panel b) of the figure, illustrates how prewhitening works.  We 
could in principle defend the choice of any of the three largest remaining 
peaks as representing real power, but we again chose the largest and most 
probable peak, and removed it.  The remaining power, shown in panel c) of 
Figure~4, is also consistent with a single periodicity.  When it too is 
removed, no power remains above noise, as shown in panel d).  Figure~5 further 
demonstrates that our solution is consistent with the original data. The top 
panel shows the power spectrum of PG~0122, while the lower panel depicts the 
power spectrum of a noise free time series, sampled at the same times as the 
data, constructed using three sine waves of the frequency, amplitude, and 
phase determined by a simultaneous least--squares solution.  Choosing 
different aliases at each stage of prewhitening might lead to equally good 
reconstructions but would require that we choose less probable peaks, leading 
in some cases to the further necessity of invoking a greater number of peaks 
to explain the same region of power.  

To within the formal error from the least-squares fit, the splitting is 
uniform, with $<$$\delta\nu$$>$~$=3.5\pm0.3$~$\mu$Hz.  This group is thus 
entirely consistent with our hypothesis that the power in PG~0122 consists 
primarily of $\ell=1$ triplet modes.

The next largest groups are those at 2220~$\mu$Hz and 2970~$\mu$Hz.  
Successive prewhitening, using the same prescription as above, revealed that 
each consists of two peaks separated by roughly twice $\delta\nu$ found in the 
previously identified triplet:  $\delta\nu=7.9\pm0.2$ and $7.4\pm0.2$ for the 
2220~$\mu$Hz and 2970~$\mu$Hz groups, respectively.  Again, in selecting peaks 
we always chose the most probable peak remaining after prewhitening to be 
``real.''  This choice is supported by identification of frequency splittings 
consistent with those in the largest group.  These two groups can be 
identified as $\ell=1$ if we allow that in each case a central ($m=0$) mode 
has not been detected.  This is highly plausible in light of the results of 
the study of PG~1159 and PG~2131; in both stars the ratio of the amplitudes of 
modes of different $m$ within a multiplet can approach an order of magnitude.  
This degree of amplitude asymmetry could easily place one or more peaks in a 
given multiplet below the level of noise. 
 
There are two more groups which show apparent multiplet structure, at 
2640~$\mu$Hz and at 2140~$\mu$Hz.  The first of the these was found to be a 
doublet with frequency splitting $\delta\nu=3.3\pm0.3$~$\mu$Hz.  Here, too, 
our simple algorithm of selecting the most likely peaks as real leads to the 
identification of a splitting consistent with every other group analyzed thus 
far.  This leaves only the group at 2140~$\mu$Hz, which we now discuss at some 
length, since it does not fit the established pattern.

The prewhitening procedure shows that the 2140~$\mu$Hz group is a doublet with 
a splitting of $5.4\pm0.3$~$\mu$Hz, a number inconsistent with identification 
of this mode with the same value of $\ell$ as the other peaks in the power 
spectrum.  However, the ratio of the splitting in the $\ell=1$ multiplets to 
that of the 2140~$\mu$Hz doublet is $0.67\pm0.04$, which is approximately the 
asymptotic ratio of $0.6$ between $\ell=1$ and $\ell=2$ rotational splittings 
(see below).  Identifying this group as $\ell=2$ explains its failure to 
conform to the splitting found in every other multiplet.  

In the final stages of writing this paper, we received a preprint of a paper 
by Vauclair et al. (1995) reporting on analysis of photometry of PG~0122 in 
October 1990.  One of the principal problems they discuss is the 
identification of $\ell$ for the 2140~$\mu$Hz multiplet. In our data two 
consecutive--$m$ modes are present, allowing us to positively identify the 
$\ell=2$ frequency splitting of 5.4~$\mu$Hz.  The two peaks present in their 
data are not consecutive in $m$.  This is not unusual behavior for the GW~Vir 
stars; they frequently show season--to--season amplitude variations.  Thus the 
$\ell=2$ splitting was not apparent in 1990, and Vauclair et al. decided that 
this mode was probably a mixture of $\ell=1$ and $\ell=2$.  When our peak list 
for this multiplet is combined with the one from Vauclair et al., the result 
is a single group of three peaks with splittings of 
$\delta\nu=5.4\pm0.3$~$\mu$Hz and 
$\delta\nu=13.9$~$\mu$Hz~=~$3\times4.6$~$\mu$Hz.  The combined list represents 
three of five possible $\ell=2$ components with a splitting of 
$\approx 5$~$\mu$Hz.

In an important respect, Vauclair et al.'s observations also reinforce our 
idea of the 2970~$\mu$Hz multiplet as an $\ell=1$ mode.  In our data, we see a 
doublet ($f=2966.2$~$\mu$Hz and 2973.6~$\mu$Hz) with the central ($m=0$) mode 
missing.  Vauclair et al. see a mode at 2970.2~$\mu$Hz, halfway between our 
two modes, providing further confirmation of the identification of the 
$\ell=1$ splitting of $3.60 \pm 0.2$~$\mu$Hz.  

At this point we have no doubt this analysis, based at every stage on the 
simplest and most likely deconvolution of individual regions of power, has 
borne out the hypothesis that the power spectrum of PG~0122 is dominated by 
$\ell=1$ modes.  A total of thirteen independent periods were identified in 
the light curve (see Table~2); all but two are members of multiplets in the 
power spectrum.  The average frequency splitting in these multiplets can now 
be used to determine the rotation rate of PG~0122. 

The frequency splitting for peaks of consecutive $m$ in each mode is set by 
the stellar rotation period $\Pi_{rot}$ according to
\begin{equation}
\delta\nu_{n,\ell} = \Pi^{-1}_{rot} \left(1 - C_{n,\ell}\right)
\end{equation}
where $C_{n,\ell}\approx(\ell(\ell+1))^{-1}$.  This splitting is not uniform 
among the $\ell=1$ modes we identified.  The minimum range, given the 
uncertainty in the frequency determinations, is 7\% from the smallest 
splitting to the largest.  This result is not unusual, however; the range in 
$\delta\nu$ among the most certain modes identified in PG~1159 is close to 
10\% (Winget et al. 1991).  The average value found in the four $\ell=1$ modes 
in the power spectrum of PG~0122 is $3.60 \pm 0.08 \mu$Hz.  This implies that 
PG~0122 rotates about its axis once every $1.61 \pm 0.04$ days.  Note that the 
identification of the peaks near 2140 $\mu$Hz as $\ell=2$ allows a 
determination of the rotation rate from that mode alone of $1.79 \pm 0.10$ 
days; this is approximately the same rotation rate derived using the triplet 
modes.

\subsection{Period Spacings} 

The most striking feature of the frequency pattern is its simplicity---very 
few modes account completely for a complex light curve.  With the exception of 
the $\ell=2$ doublet at 465~s (2140~$\mu$Hz), every multiplet is consistent 
with $\ell=1$.  We show below that, as expected from theory and previous 
experience, the $\ell=1$ modes are each separated from the others by multiples 
of a fundamental period spacing.  

The evidence for equal period spacing in the power spectrum of PG~0122 is 
summarized in Table~3.  This simple model, with $\Delta\Pi=21.2$~s, accounts 
for every $\ell=1$ mode identified in the star.  This period spacing, for 
$\ell=1$, implies a mass of $\sim0.6M_{\odot}$.  In addition, the two singlet 
modes at 570~s (1754~$\mu$Hz) and 612~s (1632~$\mu$Hz) fit the $\ell=1$ period 
spacing quite well, though of course the value of $\ell$ for these two peaks 
cannot be constrained from frequency splitting.  Finally, the singlet just 
above the surrounding power at 1430~$\mu$Hz fits well into the pattern.  
Though these singlet peaks cannot stand alone in support of the proposed 
pattern, they are accounted for by it.

The proximity of the 465~s mode to the one at 450~s implies that one of them
has a different value of $\ell$ than the other modes.  The 465~s mode fits the 
period spacing pattern far better than does the 450~s mode, which tends to 
support the identification of the 465~s mode as $\ell=1$ and the 450~s mode as 
$\ell>1$.  Unfortunately, as discussed in the previous section, the frequency 
splittings in each mode support the opposite conclusion:  $\ell=2$ for the 
465~s mode, and $\ell=1$ for the 450~s mode.  However, we prefer the 
identifications based on frequency splitting, since mode trapping can cause 
the period spacing to vary from mode to mode over a much larger range than 
will the frequency splittings ($\Delta\Pi_{n,n+1}$ varies by about 20\% of 
$<$$\Delta\Pi>$ in PG~1159, for instance).

The pulsational data are thus consistent with a pattern of $g$--mode periods
seen independently in previous observations of PG~1159 and PG~2131.  Our model 
accounts for every peak, and there is none left over.  In the only case where 
the spectrum does not conform to both equal period spacing and uniform 
frequency splitting---the 465~s doublet---theory accounts readily for the 
anomaly.  The power spectrum of PG~0122 is a set of rotationally split 
$\ell=1$ multiplets equally spaced in period, with one mode identified as 
$\ell=2$.  These results are in accord with those of the other PG~1159 stars 
previously analyzed.

These results do not agree with the conclusions reached by Vauclair et al. 
(1995).  Unfortunately, the absence of some modes robbed them of the ability 
to detect the 21.2~s period spacing.  They find 10 periods in their data.  
Seven of their ten peaks are at essentially the same frequencies in the HWN 
data we analyzed, but the amplitudes in the Vauclair et al. data averaged 3.6 
times smaller.  In particular, the triplet at 379.6~s was absent in their 
data.  As mentioned before, this behavior is not atypical of the GW~Vir stars.

The primary problem this created was a misidentification of the 465~s 
multiplet as a mixture of $\ell=1$ and $\ell=2$.  This caused Vauclair et al. 
to conclude incorrectly that the period spacing was 16.1~s.  Discarding the 
465~s group  from measurements of $\ell=1$ period spacing makes the 
observations of Vauclair et al. inadequate to determine $\Delta\Pi$ without 
ambiguity; they only have three $\ell=1$ modes, none consecutive in $n$.  We 
are fortunate to have two consecutive $\ell=1$ modes (379.6~s and 401.0~s) in 
our data.  The spacing between them (21.4~s) pointed us toward a pattern with 
an average period spacing of 21.2~s.  We are confident that $\ell=1$ modes 
found in future observations of PG~0122 will conform to this pattern and not 
to one based on 16.1~s. 

\section{The Mass of PG~0122}

The period spacing between $\ell=1$ modes constrains the global properties of 
PG~0122. To determine these properties, we computed evolutionary stellar 
models with masses between 0.56 and 0.66 M$_{\odot}$, over a range of 
effective temperatures from 140,000~K down to 70,000~K, using the stellar 
evolution code ISUEVO.  Details concerning this code can be found in Dehner 
(1995) and Dehner \& Kawaler (1995).  The helium layer thickness in all the 
models was the same as that of the best model found by Kawaler \& Bradley 
(1994) for PG~1159.  Since there are not enough consecutive modes in PG~0122 
to attempt a separate determination of the helium layer thickness, and since 
this parameter has only a very slight effect on the period spacing in the 
models anyway (Kawaler \& Bradley 1994), an order--of--magnitude estimate is 
sufficient for the present analysis. 

Using the same techniques as Kawaler \& Bradley (1994) and Kawaler et al. 
(1995), we directly compared the periods of each model to those found in the 
star.  The best fit model (in a least--squares sense) has a mass of 
$0.66M_{\odot}$, but a systematic residual exists in the fit.  Therefore we 
caution against taking this value too seriously until more careful modeling 
can be done.  The model with the best-fitting periods has an average period 
spacing of 21.6~s.  

By varying both the model mass and temperature, many models are found which do 
match the period spacing of PG~0122.  Figure~6 shows the region of the 
log(g)--log($T_{eff}$) plane occupied by the models.  A linear fit to the 
$m=0$ periods in PG~0122 yields a final spacing 
$\Delta\Pi=21.21\pm^{0.32}_{0.35}$~s.  The solid line in the figure indicates 
the interpolated position of models which satisfy this condition.  We would 
like of course to reduce the range of possibilities from a line to a point.  
This suggests a two parameter fit to the periods (of both the star and the 
models) of the form
\begin{equation}
\Pi_n = \Delta\Pi \left(n + \epsilon\right) 
\end{equation}
where both $\Delta\Pi$ and $\epsilon$ are free parameters.  This is simply the 
asymptotic period equation derived using a WKB analysis, familiar from 
standard stellar pulsation theory (c.f. Unno et al. 1989).  Though $\epsilon$ 
does not represent any obvious physical quantity, its value is set by the 
input physics and by the equilibrium stellar structure.  This is apparent 
since we found that the fit of the periods to equation [4] for every model, 
over the entire range of masses and temperatures mentioned above, gives 
essentially the $same$ value of $\epsilon$, ($\sim2.2$). This method of model 
comparison has the advantage that both $\Delta\Pi$ and $\epsilon$ are 
evaluated as an average over several modes, and therefore are affected only 
slightly by changes in the helium layer thickness, even when only a few modes 
are used (as long as the periods span several trapping cycles).  We find with 
this analysis that none of the models adequately reproduce $\epsilon$ as 
measured in the stars themselves with sufficient accuracy to distinguish them 
from one another, and therefore all models with the correct period 
spacing---regardless of the periods themselves---represent equivalent fits to 
the power spectrum of PG~0122.  Put another way, it was not possible to find a 
satisfactory and unique model which reproduced the periods and the mean 
spacing simultaneously.

We are reduced to using the spectroscopic temperature to eliminate some models 
from consideration.  The area of the H--R diagram where the correct range of 
period spacing intersects the allowed temperature region is indicated by the 
parallelogram in Figure~6.  Since this area falls outside the range of 
available models, the final mass was found by extrapolating both the 21.21~s 
line and the model masses. The mass of PG~0122 thus falls in the range 0.66 to 
0.72~$M_{\odot}$.  Until models with the same period spacing can be 
distinguished from one another (using $\epsilon$ or some other parameter), no 
independent estimate of the effective temperature (and therefore of the 
luminosity and distance) can be given.  We are preparing a paper which 
describes the details of this method of model comparison and its application 
to other PG~1159 stars.

One final point should be made concerning the power spectrum of PG~0122.  This 
is the third pre--white dwarf star, out of three studied in detail so far, 
with a period spacing near 21.5~s.  Why the mass and temperature should 
conspire, over a large range of both quantities, to give the same period 
spacing in all three stars is a mystery.  Perhaps this value for $\Delta\Pi$ 
is also a condition for pulsation?  We intend to explore the implications of 
these results for the instability strip occupied by the PG~1159 stars. 

\section{Summary and Conclusions}

Reanalysis of 1986 time--series photometry of the pulsating PG~1159 star 
PG~0122 revealed a wealth of new information about this pre--white dwarf 
star.  PG~0122 is a non--radial {\it g-}mode pulsator with several $\ell=1$ 
modes and one $\ell=2$ mode present.  Fine structure in the power spectrum
indicates that it rotates once every 1.6 days.  Comparison of the calculated 
period spacing of PG~0122 with that of stellar models implies a mass of 
0.66--0.72~$M_{\odot}$, given the spectroscopic constraint of its effective 
temperature.  

PG~0122 is similar to PG~1159 in its pulsational structure and rotation rate 
and closely resembles PG~2131 in both mass and in the quantity of modes 
observed.  We found an insufficient number of consecutive--$n$ modes with 
which to analyze possible mode--trapping in PG~0122.  Future multi--site 
observation could provide a means to study such structure and to thereby 
measure the surface helium layer mass.  This measurement is crucial, since it 
is the surface layer thickness which primarily governs the subsequent 
evolution of these stars as they become white dwarfs.  However, with global 
parameters of three pulsating pre--white dwarfs now asteroseismologically 
constrained, we have begun to establish and to understand the general 
characteristics of this important evolutionary link between the AGB stars and 
the white dwarfs.

The authors wish to acknowledge an anonymous referee for many thoughtful
suggestions and comments.  This work was supported in part by NSF Young
Investigator Award AST-9257049 to Iowa State University (MSO'B, SDK, and BTD).
Support for this work was also provided by NASA through grant number
HF-01041.01-93A from the Space Telescope Science Institute, which is operated
by the Association of Universities for Research in Astronomy, Inc., under
NASA contract NAS5-26555.

% \input{tables}
% \end{document}

\clearpage

\begin{table}
\begin{center}
\caption{Observing Log for the Archival McDonald 2.1~m
Observations of PG~0122}
\vspace*{0.5in}
\begin{tabular}{cccc}
\tableline
\tableline
                & Start Date  & Start Time & Duration \\
Run Name        &   (UT)      &    (UT)    & (h:mm:ss) \\
\tableline

RUN22 & 28 Nov 1986 & 01:47:00 & 5:58:25\\
RUN24 & 29 Nov 1986 & 01:33:50 & 2:20:40\\
RUN25 & 29 Nov 1986 & 03:54:43 & 3:44:50\\
RUN26 & 30 Nov 1986 & 01:19:00 & 2:47:40\\
RUN27 & 30 Nov 1986 & 04:08:10 & 3:14:50\\
RUN29 & 01 Dec 1986 & 02:49:30 & 5:22:30\\

\end{tabular}
\end{center}
\end{table}

\begin{table}
\begin{center}
\caption{Periodicities of PG 0122.  Numbers in parentheses
show the mean consecutive $m$ splitting if the observed
doublet is assumed to represent $m=\pm1, \ell=1$.  Commas
separate possible $m$ identifications when constraint to 
a single value was not possible. Frequency and amplitude
errors derive from a formal least squares analysis of the data.
The amplitude units are modulation amplitude, ma = $\Delta$I/I.}
\vspace*{0.5in}
\begin{tabular}{ccccccccc}
\tableline
\tableline
Period&Frequency&$\sigma_{f}$&Amplitude&$\sigma_{A}$&&&$\delta\nu$
&$\sigma_{\delta\nu}$\\
(s)&($\mu$Hz)& ($\mu$Hz)&(ma)&(ma)&$\ell$&$m$&($\mu$Hz)&($\mu$Hz)\\
\tableline 
      &        &     &     &     &  &      & & \\
612.4 & 1632.8 & 0.2 & 2.6 & 0.3 &? & ?    & &\\
      &        &     &     &     &        &  &    & \\
570.0 & 1754.4 & 0.2 & 2.5 & 0.3 &? & ?    & &\\
      &        &     &     &     &        &  &    & \\
466.4 & 2144.2 & 0.2 & 3.3 & 0.4 &2 & ?    && \vspace{-.1in}\\
      &        &     &     &     &        &      & 5.4 &0.3 \vspace{-.1in}\\
465.2 & 2149.6 & 0.2 & 2.3 & 0.4 &2 & ?    & &\\
      &        &     &     &     &        &  &    & \\
451.9 & 2213.1 & 0.2 & 4.8 & 0.4 &1 & $+1$ & &\vspace{-.1in}\\
  & & & & & & &7.9~($2\times3.95$)&0.2 \vspace{-.1in}\\
450.2 & 2221.0 & 0.1 & 6.3 & 0.4 &1 & $-1$ & &\\
      &        &     &     &     &        &  &    & \\
401.6 & 2490.0 & 0.1 & 7.3 & 0.4 &1 & $+1$ & &\vspace{-.1in}\\
      &        &     &     &     &        &      & 3.6&0.2 \vspace{-.1in}\\
401.0 & 2493.6 & 0.2 & 3.0 & 0.4 &1 & $0$  && \vspace{-.1in}\\
      &        &     &     &     &        &      & 3.3&0.2 \vspace{-.1in}\\
400.5 & 2496.9 & 0.1 &12.3 & 0.4 &1 & $-1$ && \\
      &        &     &     &     &        & &     & \\
380.1 & 2631.0 & 0.2 & 1.9 & 0.4 &1 &$+1,0$&& \vspace{-.1in}\\
      &        &     &     &     &        &      & 3.3&0.3 \vspace{-.1in}\\
379.6 & 2634.3 & 0.2 & 2.1 & 0.4 &1 &$0,-1$& &\\
      &        &     &     &     &        &  &    & \\
337.1 & 2966.2 & 0.2 & 3.1 & 0.4 &1 & $+1$ & &\vspace{-.1in}\\
  & & & & & &      & 7.4~($2\times3.70$)&0.2 \vspace{-.1in}\\
336.3 & 2973.6 & 0.1 & 5.1 & 0.4 &1 & $-1$ && \\

\end{tabular}
\end{center}
\end{table}

\begin{table}
\begin{center}
\caption{Comparison of the period spectrum ($\ell=1, m=0$) with a strict 
21.2~s equal spacing model.  An asterisk indicates that the period represents 
the calculated center of a doublet splitting.  Numbers in parentheses show the 
effects of assuming a value of $m$ other than $m=0$ for the identified peak.} 
\vspace*{0.5in}
\begin{tabular}{cccc}
\tableline
\tableline
$\Pi_{observed}$&$\Pi_{predicted}$&$\Delta$n&$\Pi_{observed}-\Pi_{predicted}$\\
($m=0$)        &($\Delta\Pi$=21.2 s)&         &           \\
 (s)     &  (s)            &        & (s)\\
\tableline 
         &                 &        &  \\
612.4($_{-1.5}^{+1.4}$)  & 613.0   & $+10$    &  $-0.6$ \\
       &                 &    &       \\
570.0($_{-1.3}^{+1.2}$)  & 570.6   & $+8$    &    $-0.6$  \\
        &                 &   &        \\
451.0*   & 443.4   & $+2$    &  $+7.6$    \\
         &         &                 &          \\
401.0  & 401.0           &  $ 0$ &   \\
         &         &                 &           \\
379.6,380.1 & 379.8             &  $-1$   & $+0.2,-0.3$  \\
         &         &                 &           \\
336.7*  & 337.4             &  $-3$  & $-0.7$   \\

\end{tabular}
\end{center}
\end{table}

\clearpage

\begin{figure}
%\plotone{fig1.ps}
\caption{The light curve of PG~0122.  The vertical axis has the units of 
modulation amplitude, ma = $\Delta$I/I.  Each panel represents a single 
night's data.}
\end{figure}

\begin{figure}
%\plotone{fig2.ps}
\caption{The power spectrum of the PG~0122 data, out to the highest frequency 
showing significant power above noise.  The power is given in units of 
$\mu$mp=mp/$10^{6}$, where mp = ($\Delta$I/I)$^{2}$.  The dashed lines 
correspond to false alarm probabilities of 30\%, 0.3\%, and 0.0001\%, 
indicating the chances that random noise will generate a peak at these power 
levels at least once in the frequency range 0 to 18,000~$\mu$Hz.}  
\end{figure}

\begin{figure}
%\plotone{fig3.ps}
\caption{The region of the power spectrum analyzed in this paper.  Note that 
the vertical scale is different in each panel.  The false alarm probabilities 
for three different power levels are again shown as in Figure~2.  For 
comparison, the spectral window shows the power spectrum of a single 
noise--free sinusoid sampled at the same times as the data.}
\end{figure}

\begin{figure}
%\plotone{fig4.ps}
\caption{Prewhitening sequence for the region of largest power in the 
spectrum; a) the original spectrum; b), c) and d) show the effects of 
subtraction in the time domain of one, two, and three sine waves (with periods 
corresponding to the peaks labeled 1--3) respectively, from the light curve.  
Note that the vertical scale is smaller in panels c) and d) than in the first 
two panels.  Frequencies, amplitudes, and phases for each subtracted mode were
obtained by simultaneous least--squares fitting to the original light curve.}
\end{figure}

\begin{figure}
%\plotone{fig5.ps}
\caption{Comparison of a portion of the power spectrum of PG~0122 to the 
spectrum of a (noise--free) light curve constructed using the three peaks 
identified by prewhitening.  The transform of the simulated light curve has 
been inverted in order to aid comparison of the two spectra by eye.}
\end{figure}

\begin{figure}
%\plotone{fig6.ps}
\caption{Logarithmic plot of surface gravity versus effective temperature 
showing evolutionary tracks based on ISUEVO.  The position occupied by PG~0122 
in this diagram is constrained by the spectroscopically determined temperature 
and by the measured period spacing.}
\end{figure}

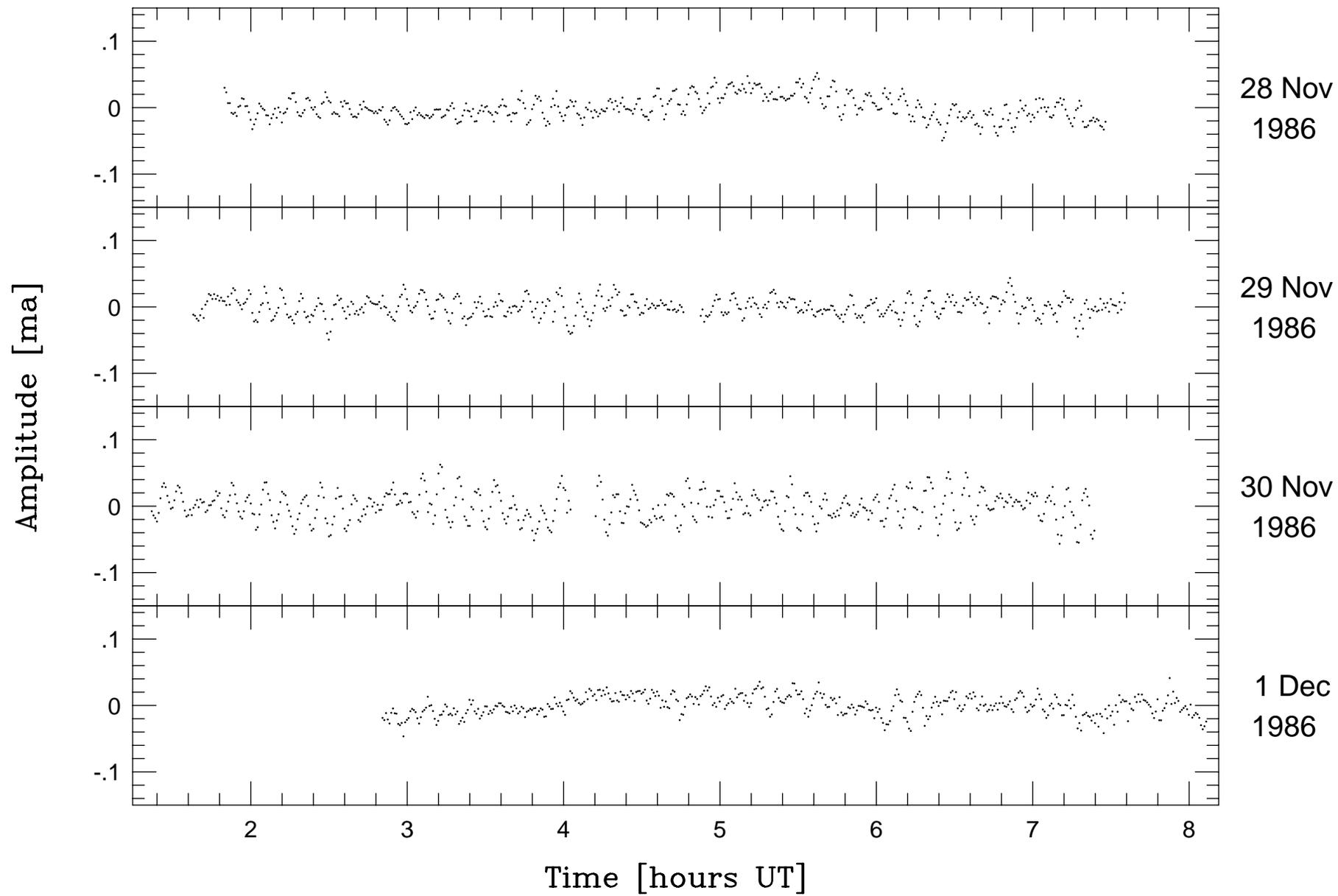
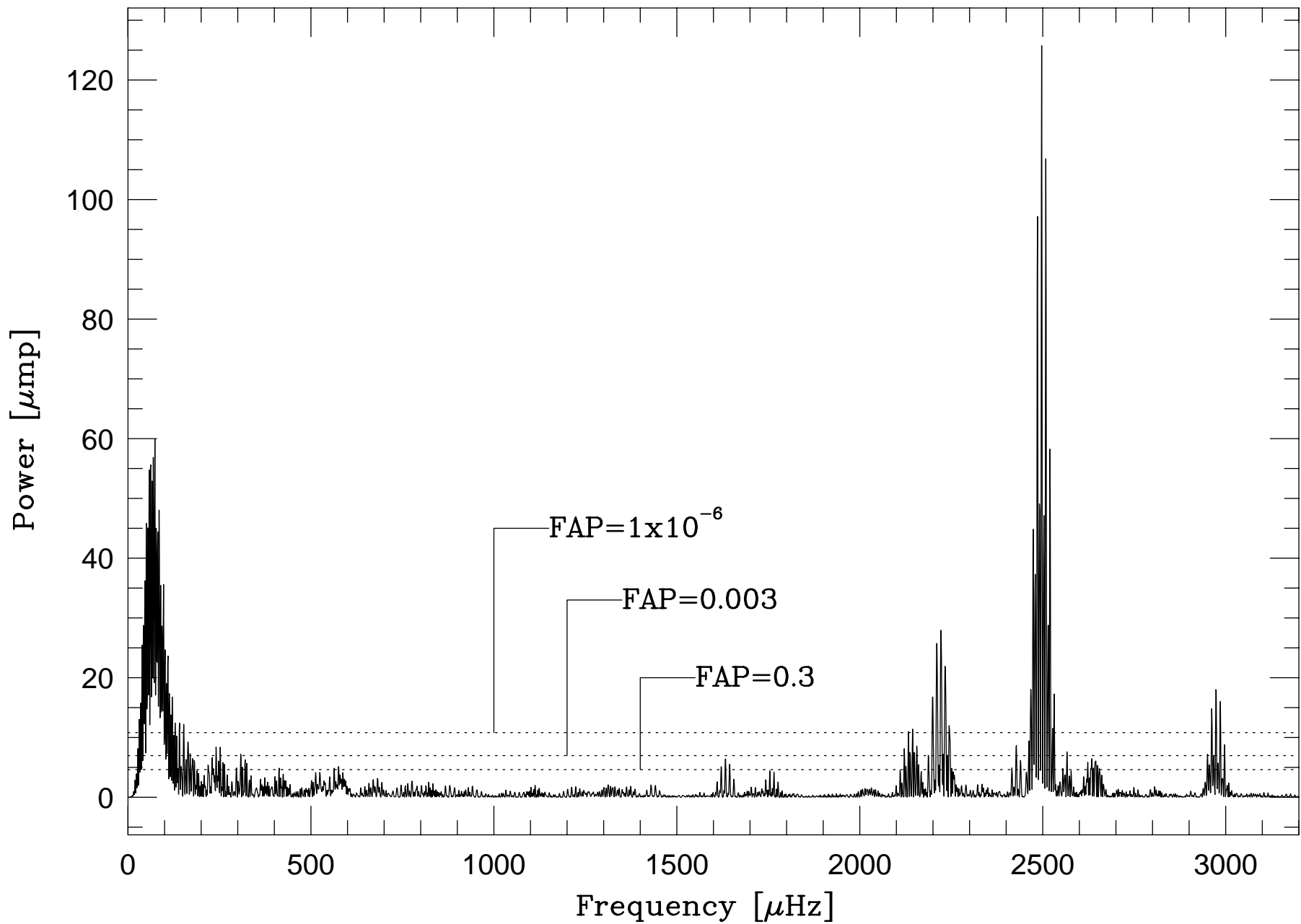
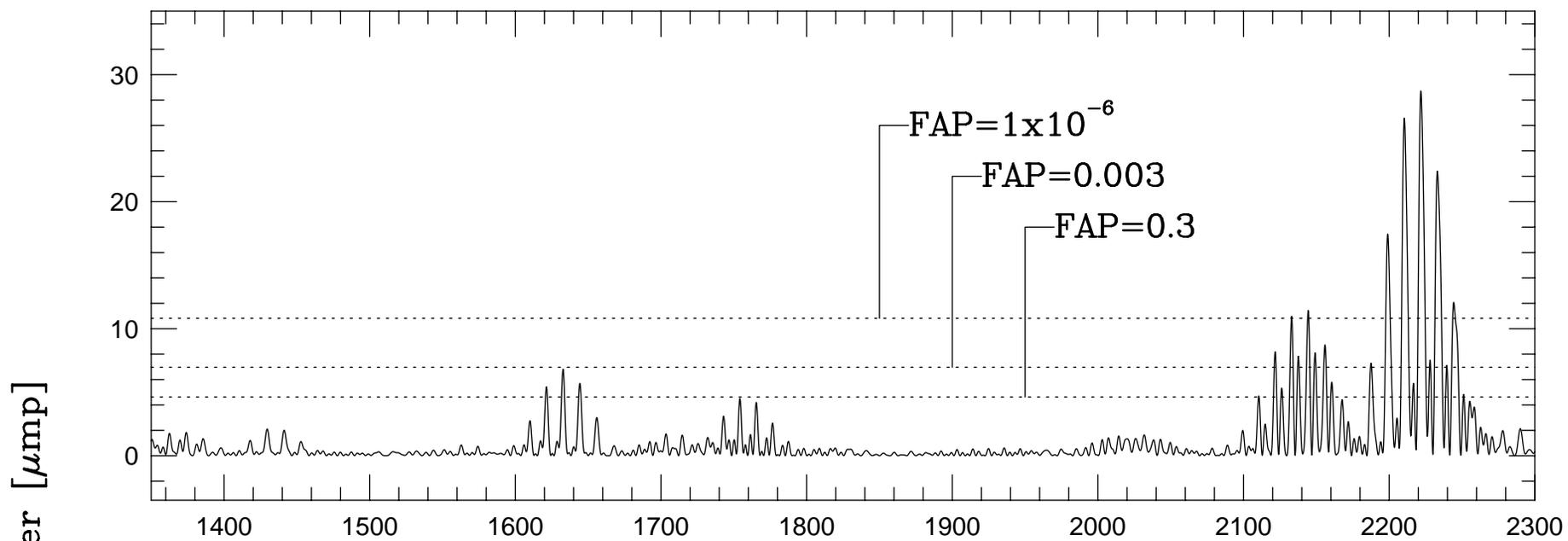
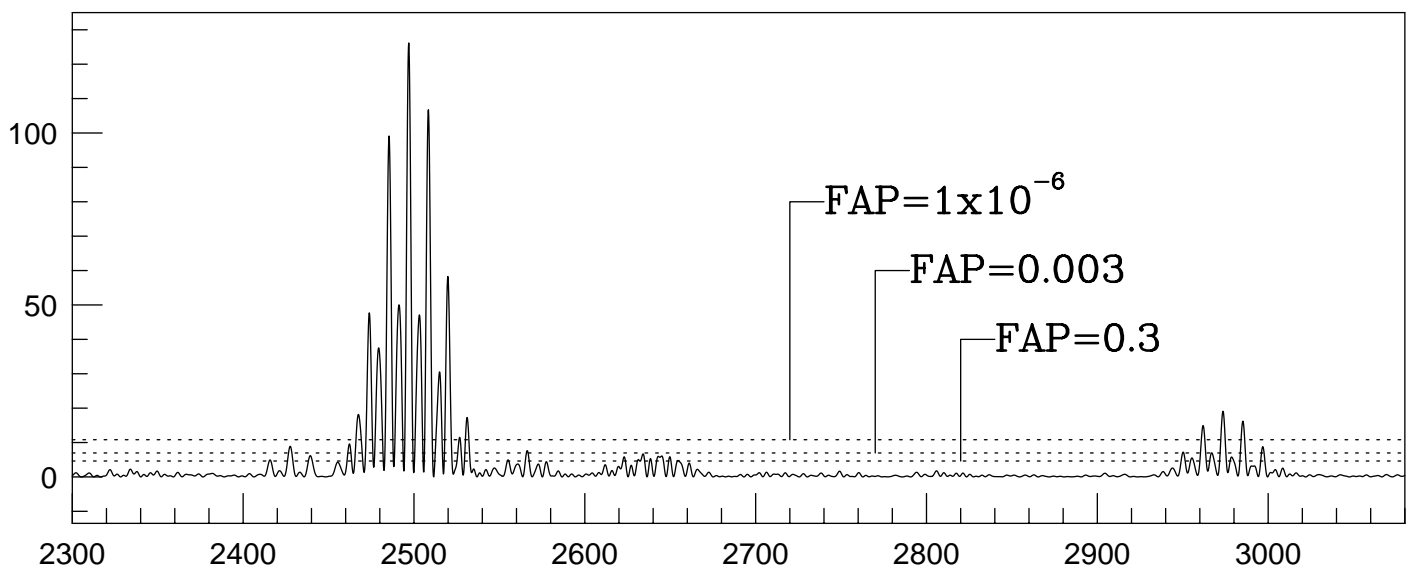
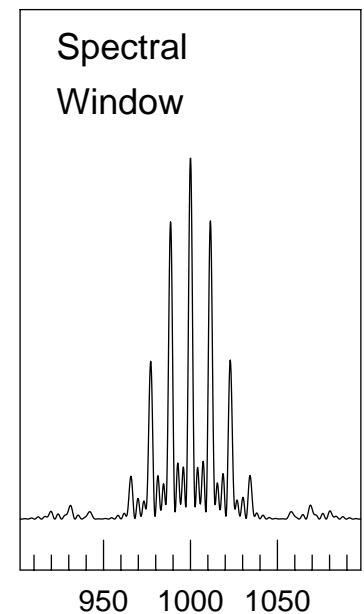
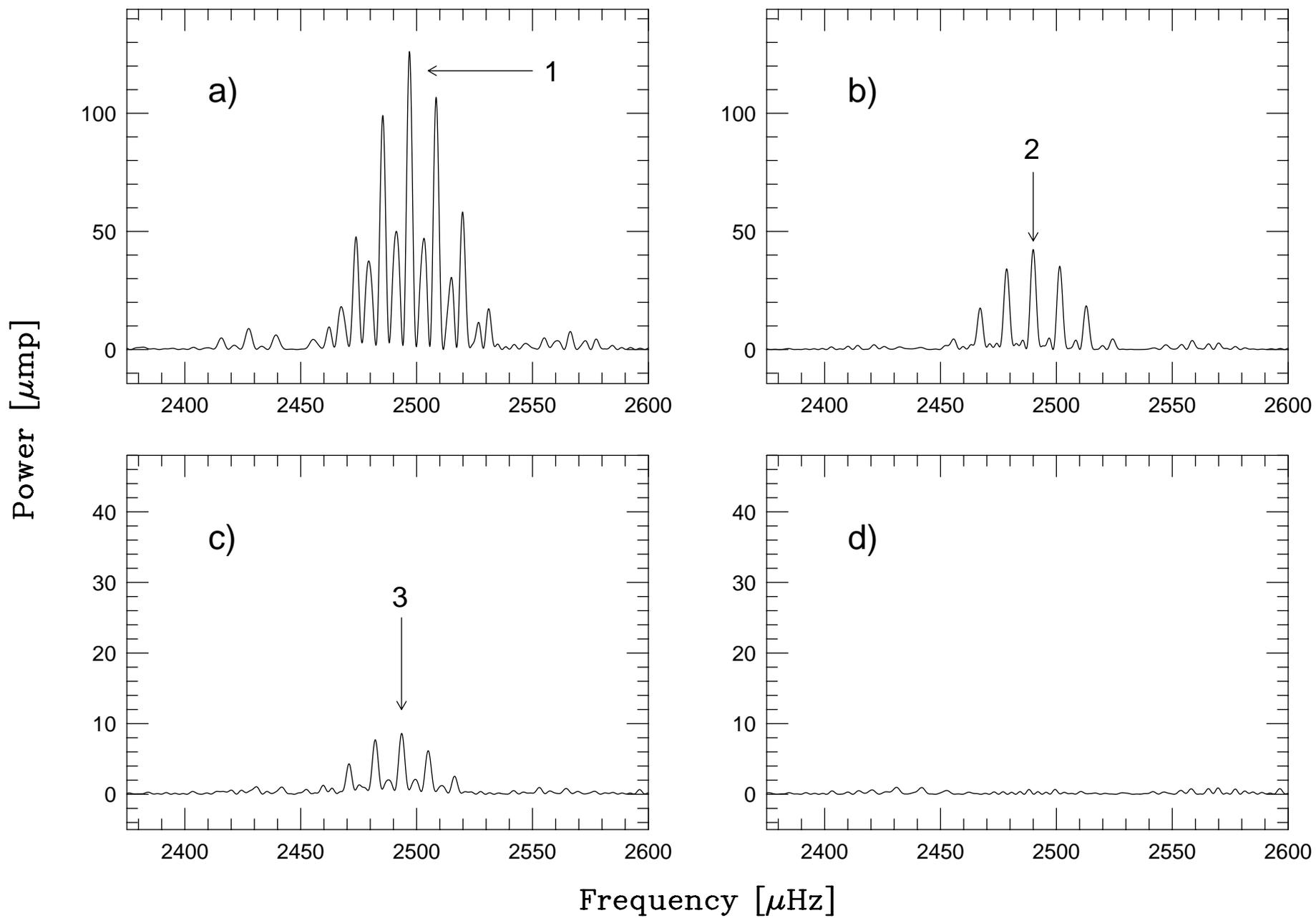
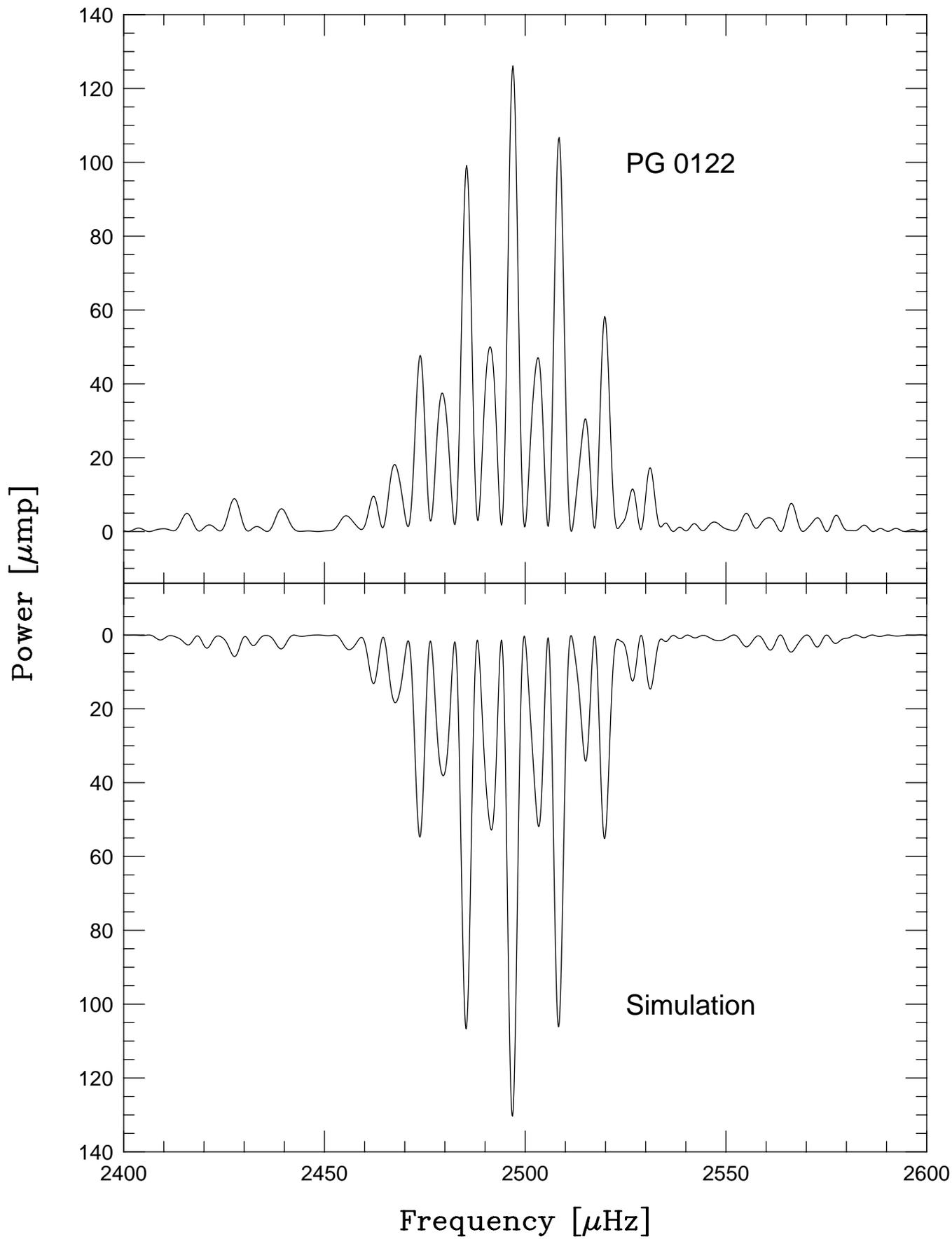
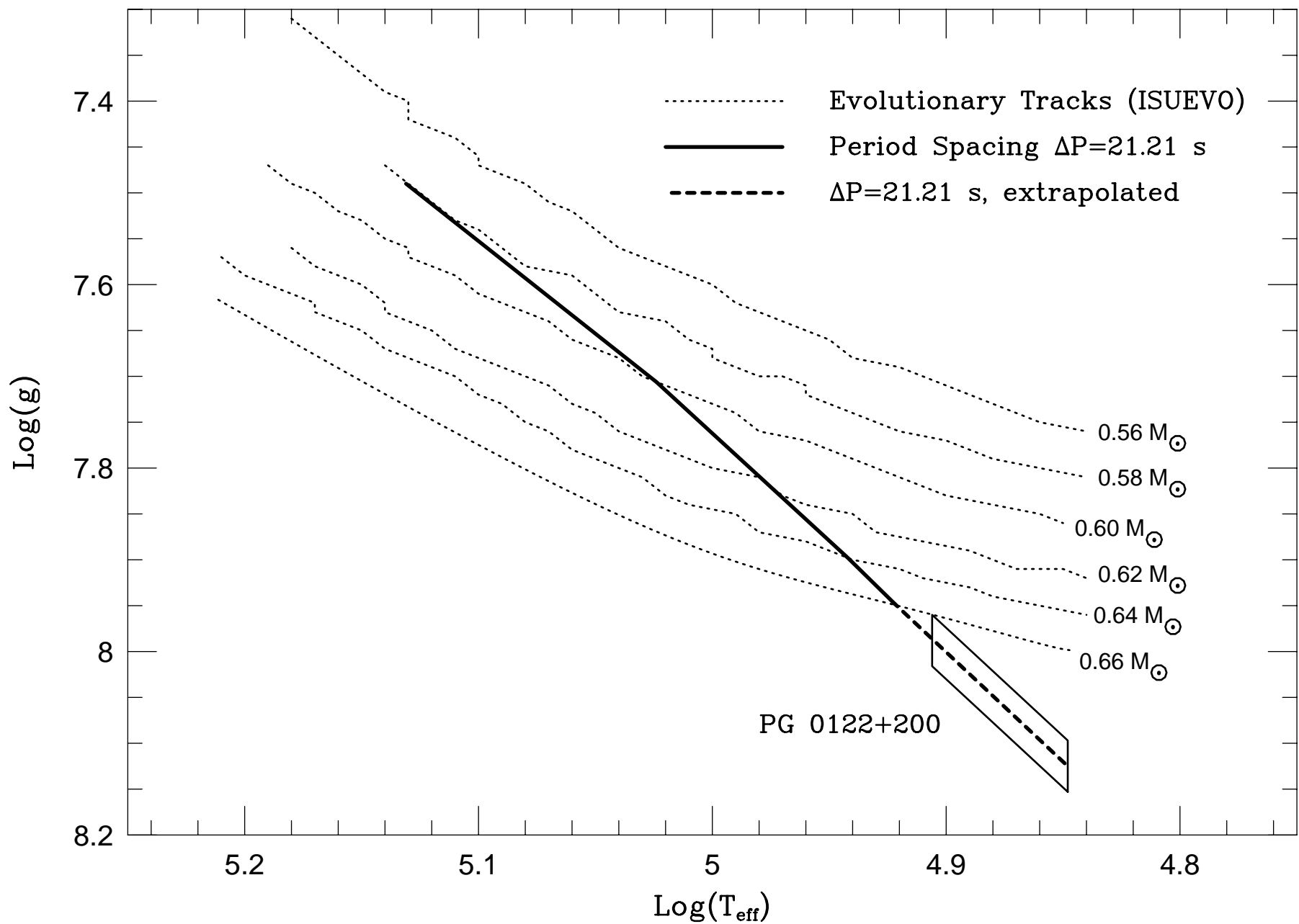
\end{document}